\documentclass[11pt]{article}

\usepackage{amsmath,amssymb}
\usepackage{graphicx,color}

\def\al{\alpha}

\def\r2{{\sqrt{2}}}

\def\Xp{X^{+}}
\def\Xm{X^{-}}
\def\Yp{Y^{+}}
\def\Ym{Y^{-}}
\def\xp{x^{+}}
\def\xm{x^{-}}
\def\yp{y^{+}}
\def\ym{y^{-}}

\begin{document}
\begin{titlepage}
\begin{flushright}
{\bf August 2006} \\ 
DAMTP-06-62\\
UT-06-16\\
hep-th/0608047 \\
\end{flushright}
\begin{centering}
\vspace{.2in}

 {\large {\bf On the Scattering of Magnon Boundstates}}

\vspace{.3in}

Heng-Yu Chen${}^{1}$, Nick Dorey${}^{1}$ and Keisuke Okamura${}^{2}$ \\
\vspace{.2 in}
${}^{1}$DAMTP, Centre for Mathematical Sciences \\ 
University of Cambridge, Wilberforce Road \\ 
Cambridge CB3 0WA, UK \\
\vspace{.2in}
and \\ 
\vspace{.1 in}
${}^{2}$Department of Physics, Faculty of Science, \\ 
University of Tokyo, Bunkyo-ku,\\
Tokyo 113-0033, Japan.\\ 
\vspace{.4in}

{\bf Abstract} \\

\end{centering}
We study the scattering of magnon boundstates in the spin-chain 
description of planar ${\cal N}=4$ SUSY Yang-Mills. Starting from the 
conjectured exact S-matrix for magnons in the $SU(2)$ sector, we
calculate the corresponding S-matrix for boundstates with an arbitrary
number of constituent magnons. The resulting expression has an
interesting analytic structure with both simple and double poles.         
We also calculate the semiclassical S-matrix for the scattering of the
corresponding excitations on the string worldsheet known as Dyonic
Giant Magnons. We find precise agreement with the magnon boundstate 
S-matrix in the limit of large 't Hooft coupling.

\end{titlepage}

\paragraph{}
The discovery of integrability in planar ${\cal N}=4$ SUSY Yang-Mills 
\cite{MZ,B} and in 
string theory on $AdS_{5}\times S^{5}$ \cite{int} has lead to considerable
progress in testing AdS/CFT duality. In particular, an 
increasingly precise correspondence between the spin-chain 
description of gauge theory 
anomalous dimensions and the world-sheet theory of the dual string 
continues to emerge \cite{BDS,AFS,Staudacher,BS,Beisert1,HM}. 
The correspondence is clearest in a 
limit where a $U(1)$ R-charges $J_{1}$ and scaling dimensions 
$\Delta$ of gauge theory operators become large.   
Following \cite{HM}\footnote{For related earlier work see
  \cite{earlier,Janik,AF}.}, the specific limit we consider is one where 
$J_{1}$, $\Delta\rightarrow \infty$ with the difference 
$E=\Delta-J_{1}$ and the
't Hooft coupling $\lambda$ held fixed. In this limit
the spin chain/string becomes infinitely long and the spectrum
consists of local excitations which propagate freely apart from
pairwise scattering. The physical content of the limiting theory 
is the spectrum of asymptotic states and their S-matrix and the main 
problem is to compare the spectrum and S-matrix which appear on
both sides of the correspondence.   
\paragraph{} 
The asymptotic spectrum of the gauge theory spin chain 
includes an infinite tower of BPS states 
labelled by a positive integer $Q$, which also carry a conserved
momentum $p$. The state $Q=1$ corresponds to 
the fundamental spin-chain excitation known as a magnon.
States with $Q>1$ correspond to boundstates of $Q$ fundamental
magnons \cite{Dorey}. Each of these states lives in a short
representation of supersymmetry and has an exact 
dispersion relation, 
\begin{equation}
E=\sqrt{Q^{2}+8g^{2}\sin^{2}\left(\frac{p}{2}\right)}
\label{Magnondispersion}
\end{equation}
where, following the convention of \cite{BDS}, we introduce a coupling 
$g$ which is related to the 
't Hooft coupling $\lambda$ by $g^{2}=\lambda/8\pi^{2}$.
For $Q>1$, this formula is a generalisation of the exact 
\cite{Beisert1} magnon dispersion relation
obtained in \cite{BDS,Staudacher,BS} (see also \cite{Dispersion}). 
\paragraph{}
In the following we will study the scattering of the BPS states
described above. The exact S-matrix for the magnons themselves 
is known up to a single overall phase \cite{Beisert1}. 
In the $SU(2)$ sector, the remaining ambiguity 
corresponds to the dressing factor first introduced in \cite{AFS, BK}. 
As we review below (see Eqn (\ref{elementarythetaAFS})), 
the dressing factor takes a very specific form as
a function of the conserved charges of the theory but still involves
an infinite number of undetermined coefficients. In an integrable
theory, the scattering of boundstates is uniquely determined by the
scattering of their constituents \cite{Zam,Karowski,Faddeev}. 
In this Letter, we will take the exact magnon S-matrix, including the 
dressing factor, as a starting point and derive 
the corresponding S-matrix for the scattering of magnon
boundstates in the $SU(2)$ sector. The resulting S-matrix has an interesting
analytic structure with simple poles corresponding to boundstate
contributions in the s- and t- channels as well as double-poles
corresponding to anomalous thresholds. The boundstate S-matrix also
includes a dressing factor which is functionally identical to the one 
appearing in the fundamental magnon S-matrix. We note that this universality
of the dressing factor is essentially equivalent to its conjectured form as a
function of the conserved charges mentioned above. 
\paragraph{} 
On the string theory side, the fundamental magnons and their
boundstates correspond to solitons of the worldsheet theory
which can be studied using semiclassical methods for $g>>1$ \cite{HM}.  
In particular, $SU(2)$ sector boundstates with values of $Q$ which
scale linearly with $g$, are identified with
classical string configurations known as ``Dyonic Giant Magnons'' 
\cite{Chen, A3, M1, SV, Bobev, Kr}. 
In string theory, $Q$ corresponds to a conserved angular momentum on
$S^{5}$ and the exact dispersion relation (\ref{Magnondispersion}) is
already obeyed at the classical level. 
These string configurations can also be mapped  
to soliton solutions of the Complex
sine-Gordon (CsG) equation using a certain reduction of the worldsheet
$\sigma$-model \cite{Chen}. In the following we
will use the CsG description to obtain a semiclassical approximation 
to the S-matrix for the Dyonic Giant Magnons. Our main result is that
this precisely matches the large-$g$ limit of the magnon boundstate
S-matrix described above. A similar comparison in the $Q=1$ case 
of fundamental magnons was performed in \cite{HM}. A new feature of
the present case is that both the dressing factor and the remaining
factor which originates from the all-loop gauge theory Bethe ansatz of
\cite{BDS} contribute at leading order 
in the $g\rightarrow \infty$ limit and therefore both parts are tested
by the comparison. While this work was being completed we learned of a
forthcoming paper \cite{Radu} with similar results.       
\paragraph{}
A generic asymptotic state in the $SU(2)$ sector has two independent
quantum numbers $p$ and $Q$. It will be convenient to use an
alternative parametrisation in terms of two complex
variables $X^{\pm}$ with,   
\begin{eqnarray}
\exp(ip)&=&\frac{X^{+}}{X^{-}}\,.\label{exponentialp}
\end{eqnarray}
The energy and charge are given respectively as, 
\begin{eqnarray}
E&=&\frac{g}{i\r2}\left[\left(X^{+}-\frac{1}{X^{+}}\right)-
\left(X^{-}-\frac{1}{X^{-}}\right)\right]\,,
\label{delta-J1}\\
Q&=&\frac{g}{i\r2}\left[\left(X^{+}+\frac{1}{X^{+}}\right)-\left(X^{-}+
\frac{1}{X^{-}}\right)\right]\,.
\label{J2}
\end{eqnarray}
If $Q$ and $E$ are regarded as free complex parameters then $X^{+}$ and
$X^{-}$ are unconstrained complex variables. 
The
condition of fixed integer charge $Q$ provides a cubic constraint
on $X^{\pm}$ which defines a complex torus \cite{Janik}. The case $Q=1$
corresponds to the fundamental magnon and the variables $X^{\pm}$
coincide with the usual spectral parameters $x^{\pm}$. We will reserve
the use of lower-case variables $x^{\pm}$, $y^{\pm}$, $\ldots$ for
this special case.  For any positive integer $Q$, physical states
with real momentum and positive energy 
are obtained by imposing the conditions $X^{-}=(X^{+})^{*}$ and $|X^{\pm}|>1$. 
\paragraph{}
It will be useful to define a rapidity variable associated with
each state\footnote{Here and in the following we will use the
  shorthand notation $f(X)$ to denote $f(X^{+},X^{-})$ defined as a
  function of two independent variables.},  
\begin{eqnarray}
U(X)\equiv
U\left(X^{+},X^{-}\right)&=&\frac{1}{2}\left[\left(X^{+}+\frac{1}{X^{+}}\right)
+\left(X^{-}+\frac{1}{X^{-}}\right)\right]
\label{UX}\,.
\end{eqnarray}
Following \cite{Arutyunov:2003rg} we also define a basis 
for the higher conserved charges carried by the BPS states,  
\begin{equation}
q_{r}(X)\equiv
q_{r}\left(\Xp,\Xm\right)=\frac{i}{r-1}\left[\frac{1}{\left(X^{+}\right)^{r-1}}
-\frac{1}{\left(X^{-}\right)^{r-1}}
\right]\,,~~~r= 2\,,3\,,\dots \label{BHighercharges}.
\end{equation} 
The rapidity and higher conserved charges, can also be given explicitly in 
terms of the momentum $p$ and charge $Q$, 
(see also \cite{M1, MinahanTalk})
\begin{eqnarray}
U(p,Q)&=&\frac{1}{g\r2}\cot\left(\frac{p}{2}\right)\sqrt{Q^{2}+8g^{2}\sin^{2}\left(\frac{p}{2}\right)}\label{UPPQ}\,,\\
q_{r}(p,Q)&=&\frac{2\sin\left(\frac{r-1}{2}p\right)}{r-1}\left(\frac{\sqrt{Q^{2}+8g^{2}\sin^{2}\left(\frac{p}{2}\right)}-Q}
{2\r2 g\sin\left(\frac{p}{2}\right)}\right)^{r-1}\,,~~~r=2\,,3\,,\dots\label{BHigherchargesPQ}.
\end{eqnarray} 
\paragraph{}
As a consequence of integrability, the asymptotic states described above undergo
factorised scattering. In other words the S-matrix for the scattering
of an arbitrary number of excitations can be written consistently as a product of
two-body factors. In the case of fundamental magnons in the SU(2)
sector, with spectral parameters $x^{\pm}$ and $y^{\pm}$ respectively,
the exact two body S-matrix 
can be written as,   
\begin{equation}
s(x,y)\equiv s\left(\xp,\xm;\yp,\ym\right) =\hat{s}(x,y)\times\sigma(x,y).\label{elementarySmatrix}
\end{equation}
The first factor,  
\begin{eqnarray}
\hat{s}(x,y)&=&\frac{x^{+}-y^{-}}{x^{-}-y^{+}}\frac{1-1/x^{+}y^{-}}{1-1/x^{-}y^{+}}=
\frac{U(x)-U(y)+\frac{\sqrt{2}i}{g}}{U(x)-U(y)-\frac{\sqrt{2}i}{g}}
\label{hatsBDS}
\end{eqnarray}
originates in the all-loop asymptotic Bethe ansatz of Beisert, Dippel
and Staudacher (BDS) \cite{BDS}. In the following we will refer to it
as the BDS factor. This factor has a pole in the
physical region of the spectral plane at the point $x^{-}=y^{+}$ which
corresponds to the formation of the $Q=2$ BPS boundstate in the
s-channel \cite{Dorey}. The second term, known as the ``dressing
factor'' \cite{AFS} corresponds to the most general
long-range integrable deformation of the Heisenberg spin chain
\cite{BK}. It can be written as,  
 \begin{eqnarray}
\sigma(x,y)&=&\exp\left(i\theta(x,y)\right)\label{singleAFS}\,.
\end{eqnarray}
where, 
\begin{equation}
\theta(x,y)=
\r2
g\sum_{r=2}^{\infty}\sum_{n=0}^{\infty}c_{r,r+1+2n}(g)[q_{r}(x)q_{r+1+2n}(y)-q_{r}(y)q_{r+1+2n}(x)]
\label{elementarythetaAFS}\,.
\end{equation}
Here $q_{r}$ are the higher charges defined in (\ref{BHighercharges})
and an infinite tower of unknown coefficients $c_{r,s}$ depending on
the coupling constant $g$ remain to be determined. 
Agreement with classical string theory uniquely determines the large-$g$ behaviour
of these coefficients as \cite{AFS},  
\begin{equation}
c_{r,s}(g)=\delta_{r+1,s}+{\mathcal{O}}\left(\frac{1}{g}\right)\label{crs}\,.
\end{equation}
More recently, the leading corrections in $1/g$ have been determined
by comparison with the one-loop corrections in the string
$\sigma$-model \cite{Hernandez:2006tk}.   
\paragraph{}
An important feature of the general form (\ref{elementarythetaAFS}) 
is that it is bilinear in the conserved charges $q_{r}$. As a
consequence, the scattering phase
$\theta(x,y)=\theta(x^{+},x^{-};y^{+},y^{-})$ is seperately odd
under the interchange of $x^{+}$ and $x^{-}$ and under the
interchange of $y^{+}$ and $y^{-}$.  
Because of unitarity, $\theta$
is also odd under the interchange $x^{\pm}\leftrightarrow y^{\pm}$.     
In fact, the general form (\ref{elementarythetaAFS}), together with
(\ref{BHighercharges}), implies that 
we can write $\theta$ in terms of a function $k(x,y)$ of two variables
satisfying $k(x,y)=-k(y,x)$ as, 
\begin{equation}
\theta(x^{+},x^{-},y^{+},y^{-})=\r2 g\left\{k(x^{+},y^{+})+k(x^{-},y^{-})
-k(x^{+},y^{-})-k(x^{-},y^{+})\right\}\label{factorizedAFS}\,.
\end{equation} 
As above $\theta$ must scale like $g$ in the strong coupling
limit. Thus the function $k$ has an expansion of the form,
\begin{equation} 
k(x,y)=k_{0}(x,y)+ \frac{1}{g}\,k_{1}(x,y)+{\mathcal{O}}\left(\frac{1}{g^{2}}\right) 
\end{equation}
and the leading term can be deduced from the leading order result
(\ref{crs}) to be \cite{AF}, 
\begin{equation}
k_{0}(x,y)=-\left[\left(x+\frac{1}{x}\right)-\left(y+\frac{1}{y}\right)\right]\log\left(1-\frac{1}{xy}\right)
\label{elemetarychi0}\,,
\end{equation} 
\paragraph{}
Let us now consider two magnon boundstates with charges
$Q_{1}\geq Q_{2}$ and momenta $p_{1}$ and $p_{2}$
respectively. Equivalently we can describe these states with
spectral parameters $X^{+}$, $X^{-}$, $Y^{+}$ and $Y^{-}$ with, 
\begin{eqnarray}
\exp(ip_{1})=\frac{X^{+}}{X^{-}}\,, & \qquad{} & \exp(ip_{2})=\frac{Y^{+}}{Y^{-}}\,,
\end{eqnarray}
where $X^{\pm}$ satisfies (\ref{J2}) with $Q=Q_{1}$ and a similar
equation holds for $Y^{\pm}$ with $Q=Q_{2}$.  
Our goal is to find the S-matrix $S(X,Y)$ describing the scattering of
these two boundstates states. In an integrable quantum theory, the S-matrix for the scattering of
boundstates is uniquely determined by the S-matrix of their
constituents. Thus, in the present case, we begin by considering the
scattering of $Q_{1}+Q_{2}$ fundamental magnons with individual
spectral parameters,   
\begin{equation}
x_{j_{1}}^{\pm}\,,~~y_{j_{2}}^{\pm}~~~{\mathrm{with}}~~j_{1}=1,\dots,Q_{1}\,,~j_{2}=1,\dots,Q_{2}\,.\label{xj1xj2}
\end{equation}
As above the spectral parameters for fundamental magnons satisfy the constraints, 
\begin{eqnarray}
&&\left(x_{j_{1}}^{+}+\frac{1}{x_{j_{1}}^{+}}\right)-\left(x_{j_{1}}^{-}+\frac{1}{x_{j_{1}}^{-}}\right)=
i\frac{\r2}{g}\,,~~j_{1}=1,\dots,Q_{1}\label{xj1id}\,,\\
&&\left(y_{j_{2}}^{+}+\frac{1}{y_{j_{2}}^{+}}\right)-\left(y_{j_{2}}^{-}+\frac{1}{y_{j_{2}}^{-}}\right)=
i\frac{\r2}{g}\,,~~j_{2}=1,\dots,Q_{2}\label{yj2id}\,.
\end{eqnarray}
\paragraph{}
By factorisability, the S-matrix for the scattering of the constituent magnons is simply a
product of two-body factors. The formation of two boundstates of
charges $Q_{1}$ and $Q_{2}$ corresponds to the pole in this
multi-particle S-matrix appearing at,  
\begin{eqnarray}
\xm_{j_{1}}=\xp_{j_{1}+1}\,,~~~j_{1}=1,\dots\, , Q_{1}-1\,,\label{polexj1}\\
\ym_{j_{2}}=\yp_{j_{2}+1}\,,~~~j_{2}=1,\dots\, , Q_{2}-1\,.\label{poleyj2}
\end{eqnarray}  
The resulting boundstate spectral parameters $X^{\pm}$ and $Y^{\pm}$ can then be
identified as:
\begin{eqnarray}
&&\Xp=\xp_{1}\,,~~~\Xm=\xm_{Q_{1}}\,,\label{XpXm}\\
&&\Yp=\yp_{1}\,,~~~\Ym=\ym_{Q_{2}}\,,\label{YpYm}
\end{eqnarray}
where it is easy to check that the appropriate constraint equation for
$X^{\pm}$ (i.e. Eqn (\ref{J2}) with $Q=Q_{1}$) is obeyed by virtue
of (\ref{xj1id}) and (\ref{polexj1}) and similarly for $Y^{\pm}$. 
Consistency of scattering in such that an integrable theory provides a
simple recipe for extracting the boundstate S-matrix: it is simply the
residue of the multi-particle scattering matrix 
of the constituent magnons at the pole specified above. This
prescription is most familar in the context of relativistic field
theories in (1+1)-dimensions \cite{Zam,Karowski}, but has also been 
applied successfully in the context of integrable spin chains
\cite{Faddeev}. In terms of
the single magnon S-matrix $s(x^{+},x^{-};y^{+},y^{-})$ given in
(\ref{elementarySmatrix}), 
the
boundstate S-matrix is,    
\begin{equation}
S\left(Q_{1},Q_{2},p_{1},p_{2}\right)=S\left(\Xp,\Xm,\Yp,\Ym\right)
=\prod_{j_{1}=1}^{Q_{1}}\prod_{j_{2}=1}^{Q_{2}}s\left(\xp_{j_{1}},\xm_{j_{2}};\yp_{j_{2}},\ym_{j_{2}}\right)\label{BSSmatrix}\,.
\end{equation}
It will be convenient to write $S$ as the product of two factors,  
\begin{equation}
S\left(Q_{1},Q_{2},p_{1},p_{2}\right)=\hat{S}\times \Sigma\label{BSSmatrix2parts}\,.
\end{equation}
Here $\hat{S}$ is the contribution coming from the BDS factor
$\hat{s}$ in the single magnon S-matrix, which is defined in
(\ref{hatsBDS}). The remaining piece $\Sigma$ originates from the
dressing factor $\sigma$ in the single magnon  S-matrix, as defined in
(\ref{elementarythetaAFS}). We will consider these two factors in
turn. 
\paragraph{}
The BDS piece of the boundstate S-matrix is straightforwardly obtained
by direct evaluation of the product (\ref{BSSmatrix}). The
corresponding calculation for the XXX Heisenberg spin chain is
reviewed in \cite{Faddeev}.
The pole
conditions (\ref{polexj1},\ref{poleyj2}) lead to numerous
cancellations between the $Q_{1}Q_{2}$ factors in the product. The
remaining factors can be conveniently presented as,  
\begin{equation}
 \hat{S}\left(Q_{1},Q_{2},p_{1},p_{2}\right)
=G\left(Q_{1}-Q_{2}\right)\left[\prod_{l=1}^{Q_{2}-1}
G\left(Q_{1}-Q_{2}+2l\right)\right]^{2}G\left(Q_{1}+Q_{2}\right)\label{BDSBSSmatrix}\,,
\end{equation}
where 
\begin{equation}
G(Q)=\frac{\Delta U+\frac{i Q}{g\r2}}{\Delta U-\frac{i Q}{g\r2}}~~~
{\mathrm{with}}~~\Delta U=U(p_{1},Q_{1})-U(p_{2},Q_{2})\label{Gfunction}\,.
\end{equation}  
\paragraph{}
The singularities of the final answer (\ref{BDSBSSmatrix}) have a
natural interpretation in terms of on-shell intermediate
states. First, the simple pole of the factor $G(Q_{1}+Q_{2})$
corresponds to the formation of a boundstate with $Q=Q_{1}+Q_{2}$ in the
s-channel. This is a direct generalisation of the $Q=2$ pole in the
S-matrix of two elementary magnons mentioned above. Similarly, the
other simple pole in $\hat{S}$, which comes from the factor
$G(Q_{1}-Q_{2})$, precisely corresponds to the exchange of a
boundstate with $Q=Q_{1}-Q_{2}>0$ in the t-channel. 
The set of $Q_{2}-1$ double poles in the boundstate S-matrix also have
a standard explanation in $(1+1)$-dimensional scattering theory
\cite{Thun}: 
they correspond to anomalous thresholds. Specifically, the positions
of the double poles are consistent with the kinematics of an 
intermediate state consisting of two on-shell boundstates with 
$Q=Q_{1}+l$ and $Q=Q_{2}-l$ respectively for $l=1,2,\ldots ,Q_{2}-1$.   
\paragraph{}
The second contribution to the boundstate scattering matrix, denoted $\Sigma$
in (\ref{BSSmatrix2parts}) comes from the dressing factors of the elementary
magnon S-matrices appearing in the product (\ref{BSSmatrix}). Here we
find an even more complete cancellation of factors appearing in the
product. In fact the final answer is simply that $\Sigma$ is identical
as function of the higher conserved charges to the fundamental magnon dressing
factor $\sigma$. Thus we have,    
\begin{equation}
\Sigma(X,Y)=\exp\left[i\theta(X,Y)\right]\,,
\end{equation}
\begin{eqnarray}
\theta(X,Y) &=&\r2
g\sum_{r=2}^{\infty}\sum_{n=0}^{\infty}c_{r,r+1+2n}(g)[q_{r}\left(X\right)q_{r+1+2n}\left(Y\right)-
q_{r}\left(Y\right)q_{r+1+2n}\left(X\right)]\,,  
\label{exactThetaAFS}
\end{eqnarray}
where the coefficient $c_{r,s}$ are the same as those appearing in
(\ref{elementarythetaAFS}). Thus the factor $\Sigma$ appearing in the
boundstate S-matrix is equal to a universal function of the higher
conserved charges. Equivalently we have 
\begin{equation}
\theta(X,Y)=\theta(X^{+},X^{-};Y^{+},Y^{-})=\r2 g\left\{k(X^{+},Y^{+})+k(X^{-},Y^{-})
-k(X^{+},Y^{-})-k(X^{-},Y^{+})\right\}\label{factorizedAFS2}\,,
\end{equation} 
where $k(X,Y)$ is the same function appearing in
(\ref{factorizedAFS}). As in the case of the single magnon S-matrix
our knowledge of this function (or, equivalently of the coefficients $c_{r,s}$) is limited to the first two orders in
the strong coupling expansion. As mentioned above, the general form
(\ref{elementarythetaAFS}) for the dressing factor originally arose as
the most general integrable long-range deformation of the Heisenberg
spin chain. In the present context it is interesting to note that it
is essentially equivalent to the condition that the dressing factor
should be the same universal function of the conserved charges for all 
BPS states in the theory. Indeed one could start by imposing this
universality as a requirement and, after also taking account of unitarity and
parity invariance, one would immediately be lead to the general form
(\ref{elementarythetaAFS}).   
\paragraph{}
So far we have been considering the exact analytic expressions for the
boundstate S-matrix. To compare our results with those of
semiclassical string theory we need
to take the strong coupling limit $g\rightarrow \infty$. As discussed
in \cite{Chen}, the natural limit to take is one where the charges
$Q_{1}$ and $Q_{2}$ also scale linearly with $g$. As a consequence, 
both terms under the square root in the dispersion relation 
(\ref{Magnondispersion}) scale like $g^{2}$ and thus the energy $E$
has the appropriate coupling dependence for a semiclassical string
state. Conveniently, the spectral parameters $X^{\pm}$ and $Y^{\pm}$ 
for boundstates with $Q=Q_{1}$ and $Q=Q_{2}$ respectively remain fixed
in this limit.  
Our next goal is to calculate the leading asymptotics of the
boundstate S-matrix as a function of the spectral parameters. 
As above we consider the two factors $\hat{S}$ and
$\Sigma$ appearing in (\ref{BSSmatrix2parts}) in turn.   
\paragraph{} 
To take the strong coupling limit of $\hat{S}$, we begin by
exponentiating the product appearing in (\ref{BDSBSSmatrix}) to obtain
a sum in the exponent. As $g\rightarrow \infty$ this sum goes over to
an integral, with the integration limits depending only on the sum and
the difference between the charges $Q_{1}$ and $Q_{2}$. 
Interestingly, the leading contribution to $\hat{S}$ has the same
general form (\ref{factorizedAFS2}) as that of the dressing factor. 
In particular, the final result can be given as,  
\begin{equation}
\hat{S}\left(X, Y\right) \sim \exp\left(i\hat{\theta}\left(X, Y \right)\right)\label{SmatrixBSb}\,,
\end{equation}
where 
\begin{equation}
\hat{\theta}\left(X^{+},X^{-},Y^{+},Y^{-}\right)
=\r2 g\left[\hat{k}\left(X^{+},Y^{+}\right)+\hat{k}\left(X^{-},Y^{-}\right)
-\hat{k}\left(X^{+},Y^{-}\right)-\hat{k}\left(X^{-},Y^{+}\right)\right]\,.\label{Deltatotalb}
\end{equation}
Here the function $\hat{k}$ is given by, 
\begin{equation}
\hat{k}\left(X,Y\right)=
\left[\left(X+\frac{1}{X}\right)-\left(Y+\frac{1}{Y}\right)\right]\log\left[(X-Y)\left(1-\frac{1}{XY}\right)\right]
\label{BDS2}\,.
\end{equation}
\paragraph{}
The strong-coupling limit of the dressing factor $\Sigma$ is simply
given by replacing the function $k(X,Y)$ appearing in
(\ref{factorizedAFS2}) by the function $k_{0}(X,Y)$ given in
(\ref{elemetarychi0}). Collecting the results for the two factors we
find the final result for the strong coupling limit of the boundstate
S-matrix can be given as, 
\begin{equation}
S\left(X, Y\right) \sim \exp\left(i\Theta\left(X, Y \right)\right)\label{SmatrixBS}\,,
\end{equation}
where 
\begin{equation}
\Theta\left(X^{+},X^{-},Y^{+},Y^{-}\right)
=\r2 g\left[K\left(X^{+},Y^{+}\right)+K\left(X^{-},Y^{-}\right)
-K\left(X^{+},Y^{-}\right)-K\left(X^{-},Y^{+}\right)\right]\,.\label{Deltatotal}
\end{equation}
Here the function $K(X,Y)$ is given by, 
\begin{equation}
K\left(X,Y\right)=\hat{k}\left(X,Y\right)+k_{0}\left(X,Y\right)
=\left[\left(X+\frac{1}{X}\right)-\left(Y+\frac{1}{Y}\right)\right]\log\left(X-Y\right)\,.\label{Kfunction}
\end{equation}
\paragraph{}
In our previous paper \cite{Chen} we showed that the magnon boundstates described
above appear in string theory on $AdS_{5}\times S^{5}$ as
classical solitons of the worldsheet action. For states in the $SU(2)$
sector we may restrict our attention to strings moving on an
$\mathbb{R}\times S^{3}$ subspace of $AdS_{5}\times S^{5}$. The
corresponding equations of motion together with the Virasoro
constraint can be mapped onto the complex sine-Gordon (CsG) equation. 
Under this equivalence, the classical string solution corresponding to a
magnon boundstate of charge $Q$ and momentum $p$ is mapped to a
certain one-soliton solution of the CsG equation. The soliton in
question has two parameters: a rapidity\footnote{Not to be confused
  with the magnon rapidity $U(X)$ introduced above.} $\theta$ and an
additional rotation parameter $\alpha$. The dictionary between these
parameters and the conserved quantities $E$, $Q$ and $p$ is,         
\begin{eqnarray}
E&=&2\r2 g\frac{\cos\left(\al \right)\cosh\left(\theta
  \right)}{\cos^{2}\left(\al \right)+\sinh^{2}\left(\theta \right)}\label{Eithetaalpha}\,,\\
Q &=&2\r2 g\frac{\cos\left(\al \right)\sin\left(\alpha
  \right)}{\cos^{2}\left(\al \right)+\sinh^{2}\left(\theta \right)}
\label{Qithetaalpha}\,,
\end{eqnarray}
and  
\begin{equation}
\cot\left(\frac{p}{2}\right)=\frac{\sinh\left(\theta
  \right)}{\cos\left(\alpha \right)}\,.
\end{equation}
\paragraph{}
The CsG equation is completely integrable and has multi-soliton
scattering solution which can be constructed explicitly via inverse
scattering \cite{MdV} or by the Hirota method \cite{Get}. The only effect
of scattering is to induce a time delay for each soliton relative to
free motion. For two solitons with rapidities $\theta_{1}$ and
$\theta_{2}$ and rotation parameters $\alpha_{1}$ and $\alpha_{2}$ 
the COM frame is defined by the condition, 
$\cos\left(\alpha_{1}\right)\sinh\left(\theta_{1}\right)=-\cos\left(\alpha_{2}\right)\sinh\left(\theta_{2}\right)$. 
In this Lorentz frame the two solitons experience an equal time delay $\Delta
T_{1}=\Delta T_{2}=\Delta T_{{\mathrm{COM}}}$
with \cite{Dorey:1994mg}, 
\begin{equation}
\Delta T_{{\mathrm{COM}}}=\frac{1}{\cos\left(\al_{1}\right)
\sinh\left(\theta_{1}\right)}\log F\left(\Delta\theta,\Delta\alpha,\bar{\alpha}\right)\,,\label{CsGtimedelay}
\end{equation}  
where we define $\Delta\theta=(\theta_{1}-\theta_{2})/2$,
$\Delta\alpha=(\al_{1}-\al_{2})/2$ and $\bar{\alpha}=
(\al_{1}+\al_{2})/2$ and the function $F$ is given by, 
\begin{equation}
F\left(\Delta\theta,\Delta\alpha,\bar{\alpha}\right)
=\frac{\sinh\left(\Delta\theta+i\Delta\al\right)\sinh\left(\Delta\theta-i\Delta\al\right)}
{\cosh\left(\Delta\theta+i\bar{\al}\right)\cosh\left(\Delta\theta-i\bar{\al}\right)}\label{Ffunction}\,.
\end{equation}
Time delays due to multiple soliton scattering are simply given by the sum
of the delays experienced in each two-body collision. This is a
consequence of integrability, and is a
classical analog of the factorisability of the S-matrix. Indeed, the
time delays determine the semiclassical approximation to the worldsheet S-matrix
$S_{\rm string}=\exp(i\Theta_{\rm string})$. In particular, if we
express the S-matrix as a function of the energies $E_{1}$ and $E_{2}$
of the two excitations and their charges $Q_{1}$ and $Q_{2}$ we have
\cite{Jackiw:1975im},      
\begin{equation}
\Delta T_{1}=\frac{\partial \Theta_{\rm string}}{\partial E_{1}}\,,
~~~\Delta T_{2}=\frac{\partial \Theta_{\rm string}}{\partial E_{2}}\,.\label{DeltaT1T2b}
\end{equation}\
\paragraph{}
Our aim here is to compare $S_{\rm string}$ with the semiclassical
limit of the magnon boundstate S-matrix computed above. 
Equivalently we can use the boundstate S-matrix to compute the time delay in
boundstate scattering directly and compare with the COM frame
expression for $\Delta T_{1}$ and $\Delta T_{2}$ presented in
(\ref{CsGtimedelay}) above. To do so, one has to first express
$\Theta \left(X^{+},X^{-},Y^{+},Y^{-}\right)$ in terms of the charges
$Q_{1}, Q_{2}$ and the energies  
$E_{1}, E_{2}$ using the relations:
\begin{eqnarray}
E_{1}&=&\frac{g}{i\r2}\left[\left(X^{+}-\frac{1}{X^{+}}\right)-\left(X^{-}-\frac{1}{X^{-}}\right)\right]\,,\label{E1}\\
E_{2}&=&\frac{g}{i\r2}\left[\left(Y^{+}-\frac{1}{Y^{+}}\right)-\left(Y^{-}-\frac{1}{Y^{-}}\right)\right]\,,\label{E2}\\
Q_{1}&=&\frac{g}{i\r2}\left[\left(X^{+}+\frac{1}{X^{+}}\right)-\left(X^{-}+\frac{1}{X^{-}}\right)\right]\,,\label{Q1}\\
Q_{2}&=&\frac{g}{i\r2}\left[\left(Y^{+}+\frac{1}{Y^{+}}\right)-\left(Y^{-}+\frac{1}{Y^{-}}\right)\right]\,.\label{Q2}
\end{eqnarray} 
We then define, 
\begin{equation}
\Delta {\tau}_{1}=\frac{\partial \Theta}{\partial E_{1}}\,,
~~~\Delta {\tau}_{2}=\frac{\partial \Theta}{\partial E_{2}}\,,\label{DeltaT1T2}
\end{equation} 
while keeping the charges $Q_{1}$ and $Q_{2}$ fixed. 
Here we present the results of explicit differentiations exclusively in terms of spectral parameters:
\begin{eqnarray}
\Delta {\tau}_{1}&=&i\frac{\left(\left(X^{+}\right)^{2}-1\right)\left(\left(X^{-}\right)^{2}-1\right)}{\left((\Xp)^{2}-(\Xm)^{2}\right)}\log\left(\frac{\Xp-\Yp}{\Xp-\Ym}\frac{\Xm-\Ym}{\Xm-\Yp}\right)
-\left(\frac{1}{\Yp}-\frac{1}{\Ym}\right)\frac{\left(\Xp\Xm+1\right)}{\left(\Xp+\Xm\right)}
\,,\label{DeltaT1exp}\nonumber\\
\\
\Delta {\tau}_{2}&=&-i\frac{\left(\left(Y^{+}\right)^{2}-1\right)\left(\left(Y^{-}\right)^{2}-1\right)}{\left((\Yp)^{2}-(\Ym)^{2}\right)}\log\left(\frac{\Xp-\Yp}{\Xp-\Ym}\frac{\Xm-\Ym}{\Xm-\Yp}\right)
+\left(\frac{1}{\Xp}-\frac{1}{\Xm}\right)\frac{\left(\Yp\Ym+1\right)}{\left(\Yp+\Ym\right)}
\,.\label{DeltaT2exp}\nonumber\\
\end{eqnarray} 
\paragraph{}
All that remains is to compare with the CsG time delays. Combining the
identities (\ref{E1})-(\ref{Q2}) and the relations
(\ref{Eithetaalpha}) 
and (\ref{Qithetaalpha}), one can express the spectral parameters of
the magnon boundstates in terms of\footnote{In obtaining these
  expression one needs to solve quadratic equations. The appropriate
  root of the quadratic is selected by demanding that the
  corresponding state has positive
  energy.} $\theta_{i}$ and $\alpha_{i}$,
\begin{eqnarray}
X^{\pm}&=&\coth\left(\frac{\theta_{1}}{2}\pm i\left(\frac{\al_{1}}{2}-\frac{\pi}{4}\right)\right)\,,\label{Xpm}\\
Y^{\pm}&=&\coth\left(\frac{\theta_{2}}{2}\pm i\left(\frac{\al_{2}}{2}-\frac{\pi}{4}\right)\right)\,.\label{Ypm}
\end{eqnarray}
These expressions in turn yield
\begin{equation}
F\left(\Delta\theta,\Delta\alpha,\bar{\alpha}\right)\equiv
F\left(\Xp,\Xm,\Yp,\Ym\right)=\frac{\left(\Xp-\Yp\right)\left(\Xm-\Ym\right)}
{\left(\Xp-\Ym\right)\left(\Xm-\Yp\right)}\label{Fexplicit}\,.
\end{equation}
Now comparing (\ref{CsGtimedelay}) with (\ref{DeltaT1exp}) and
(\ref{DeltaT2exp}), taking into account the COM frame condition, we
can see that the time-delays for boundstate scattering agree with 
those of CsG solitons up to a specific non-logarithmic term, 
\begin{equation}
\Delta \tau_{1}=
\Delta T_{1}+\left(\frac{1}{\Yp}-\frac{1}{\Ym}\right)\frac{\left(\Xp\Xm+1\right)}{\left(\Xp+\Xm\right)}\,.
\end{equation}
Upon integration with respect to $E_{1}$, we can obtain the relation between the scattering phases:
\begin{equation}
\Theta_{\rm string}=\Theta\left(X^{+},X^{-},Y^{+},Y^{-}\right)+\left(E_{2}-Q_{2}\right)p_{1}\,.\label{phaserelation}
\end{equation}
The non-logarithmic term in (\ref{DeltaT1exp}) integrates up to give
the difference term above that is a direct generalisation of the one in eqn. (3.33)
of \cite{HM}. As in that case, the difference can be accounted for by
taking into account the different effective length of the excitation on the both sides of
the correspondence and is irrelevant for the Bethe ansatz equations. 
One can also check that the expressions in (\ref{DeltaT1exp}) and
(\ref{DeltaT2exp}) correctly satisfy $\Delta \tau_{1}=\Delta \tau_{2}$
in the COM frame. 
\paragraph{}
The authors acknowledge useful discussions with Gleb Arutyunov, 
Diego Hofman, Juan Maldacena and Radu Roiban. 
We are grateful to Radu Roiban for
providing us with a preliminary version of \cite{Radu}. 
The authors would also like to thank Rui
Fernando Lima Matos for collaboration on some of the work presented
here. HYC is supported by a Benefactors' scholarship from St.\,John's
College, Cambridge. ND is supported by a PPARC Senior Fellowship. 
HYC would like to thank National Taiwan University, Physics Department 
for the hospitality, where part of the work was completed.  KO is very
grateful to University of Cambridge, 
Centre for Mathematical Sciences, where part of the work was done, 
for its warm hospitality.


\end{document}